\documentclass[pre,twocolumn]{revtex4}
\usepackage[english]{babel}
\usepackage{amsmath}
\usepackage{amssymb}
\usepackage{epsfig}

\begin{document}

\title{Optimal dynamics of a spherical squirmer in Eulerian description}
\author{V.~P. Ruban}
\email{ruban@itp.ac.ru}
\affiliation{Landau Institute for Theoretical Physics RAS,
Chernogolovka, Moscow region, 14243 Russia} 
\date{\today}

\begin{abstract}
The problem of optimization of a cycle of tangential deformations 
of the surface of a spherical object (microsquirmer) self-propelling 
in a viscous fluid at low Reynolds numbers is represented in a 
noncanonical Hamiltonian form. The evolution system of equations for 
the coefficients of expansion of the surface velocity in the associated 
Legendre polynomials $P^1_n(\cos\theta)$ is obtained. The system is 
quadratically nonlinear, but it is integrable in the three-mode 
approximation. This allows a theoretical interpretation of numerical 
results previously obtained for this problem.
\end{abstract}

\maketitle

\section*{Introduction}

The hydrodynamics of swimming microorganisms
has currently become a separate field of research at the
interface between biology and mechanics (see review
[1] and numerous references therein). In addition, this
field is of interest for the creation of medical nanorobots. 
The aim of hydrodynamics in this application is
the quantitative description of motion of a fluid and
active objects in it. For their motion, microorganisms
involve diverse tools and skills such as flagella, cilia,
and surface deformations. The decisive simplification
of the theory can be achieved because fluids flow at a
very low Reynolds numbers [2], when effects of inertia
are insignificant as compared to viscosity (Stokes
regime). Consequently, the velocity field almost
instantaneously and unambiguously responds to any
change in the shape of a body, shifting it in space. It is
important that the displacement of an object periodically 
varying its state can be nonzero in a cycle only in
the presence of a ``loop'' in the space of parameters
characterizing the shape of the body. Therefore, the
number of such time-dependent parameters, i.e.,
internal degrees of freedom, should be no less than
two. A number of proposed simplified models allowed
the comprehensive study of the mechanics of viscous
swimming. In particular, a ``three-sphere swimmer''
with two arms with a variable length is well known [3-6]. 
Other models, including much more complex
ones, were also studied (see, e.g., [2, 7-10] and references 
therein). Another type of swimming occurs
through tangential deformations of the surface of a
body without changing its geometric shape [11-15].
Such a model microorganism is called squirmer. The
simplest squirmer is spherical and allows a simple
exact solution of the problem in terms of associated
Legendre polynomials [11, 12]. More precisely, if the
tangential velocity field in the coordinate system 
associated with the sphere is expanded in the series
\begin{equation}
u_\theta(\theta,t)=\sin\theta \sum_{n=1}^{\infty}
\frac{a_n(t) P'_n(\cos\theta)}{n(n+1)},
\end{equation}
the velocity of the squirmer along its axis of symmetry
with respect to the fluid at rest at infinity is $U_{\rm sq}(t)=a_1(t)/3$.
Since the surface velocity $u_\theta(\theta,t)$ is
determined by the motion of certain Lagrangian
markers $\theta_0$,
\begin{equation}
u_\theta(\theta,t)=\frac{\partial \theta(\theta_0,t)}{\partial t}
\Big|_{\theta_0=\theta_0(\theta,t)}.
\end{equation}
The time-periodic (bijective) mapping $\theta(\theta_0,t)$ 
specifies the ``loop'' in the configuration space of the
squirmer that is responsible for its locomotion.

Here, it is reasonable to immediately estimate the
energy efficiency of any given cycle. If only the
mechanical energy dissipation in the surrounding
fluid is taken into account and internal energy consumptions 
are disregarded, the instantaneous dissipation 
rate is given by the expression
\begin{equation}
Q\propto\sum_n\frac{D_n}{2n+1}a_n^2,
\end{equation}
Here, $D_1=1$ and the other coefficients are specified by
the formula [11, 12]
\begin{equation}
D_n=\frac{2n+1}{n(n+1)}, \qquad n\geq 2.
\end{equation}
Below, the general case of arbitrary coefficients $D_n>0$
is considered with particular numerical examples.

It is convenient to reformulate the problem of optimization 
of the cycle in other variables [16]. Let $x=\cos\theta$
be a Lagrangian mapping of the surface. Then, the corresponding 
velocity field has the form
\begin{equation}
v(x,t)=-\sin\theta u_\theta=\sum_{n=1}^{\infty}a_n(t) 
\Big(\frac{d}{dx}\Big)^{-1} P_n(x),
\label{v}
\end{equation}
where
\begin{equation}
\Big(\frac{d}{dx}\Big)^{-1} P_n(x)=-\frac{(1-x^2)P'_n(x)}{n(n+1)}.
\label{d_minus_P}
\end{equation}
and
\begin{equation}
a_n(t)=-\frac{(2n+1)}{2}\int P'_n(x) v(x,t)dx.
\end{equation}
Correspondingly,
\begin{equation}
U_{\rm sq}(t)=\frac{a_1(t)}{3}=-\frac{1}{2}\int v(x,t) dx.
\label{Usq}
\end{equation}

The optimal cycle should ensure the maximum displacement 
of the sphere in period at limited energy
consumption. Such a cycle was found numerically in
[16] in terms of the mapping $x(x_0,t)$ by the gradient
maximization of the ratio $\langle a_1\rangle ^2/\langle Q\rangle$, 
where the angular
brackets stand for averaging over the period. However,
the time dependences of the coefficients $a_n(t)$ were
obtained indirectly already after solving the equation
of motion for $x(x_0,t)$ . The interpretation of higher
modes was rather difficult. In general, the analytical
analysis of the problem remains incomplete.

The aim of this work is to supplement [16] by the
theoretical analysis of the problem of optimization of
a spherical squirmer taking into account the existing
symmetry of redefinition of Lagrangian markers
$x_0=x_0(\xi_0)$. As will be shown below, the system of
equations of motion for the problem can be obtained
directly in terms of $a_n(t)$. Furthermore, these differential 
equations are only of the first order on time and are resolved
with respect to the time derivative. Moreover, the
three-mode approximation is integrable, thus completely 
revealing the structure of the corresponding solution.

\section*{Hamiltonian mechanics of optimization}

It is convenient to use standard conservative
dynamics. In view of Eq. (8), the most efficient cycle
of the squirmer should ensure the minimum of the
``action'' functional $A=\int L dt$ for the Lagrangian
\begin{equation}
L[v]=\int_{-1}^{1}\Big(\frac{1}{2}v\hat D v +\lambda v\Big)dx, 
\,\mbox{ where  } 
v=-\frac{\partial x_0/\partial t}{\partial x_0/\partial x},
\end{equation}
$\lambda$ is an indefinite Lagrange multiplier, and $\hat D$ is
the operator specified as
\begin{equation}
\hat D v =-\sum_{k=1}^{\infty} D_k a_k(t)  P'_k(x).
\label{Dv}
\end{equation}
The variation of the action with respect to $\delta x_0(x,t)$ 
easily gives the corresponding Euler-Lagrange
dynamic equation. Simple algebra reduces this equation to the equation
\begin{equation}
\frac{\partial p}{\partial t}=
-\frac{\partial}{\partial x}(vp)-p\frac{\partial v}{\partial x},
\label{p_t}
\end{equation}
where $p(x,t)=\delta L[v]/\delta v$ is the canonical momentum.
In our case,
\begin{equation}
p=\hat D v +\lambda.
\end{equation}
It is noteworthy that the Lagrangian mapping itself is
absent here and Eq. (11) is of the first order in the time derivative 
rather than of the second order. This is due to the mentioned
symmetry of the redefinition when passing from the
Lagrangian to Eulerian description.

Equation (11) has a noncanonical Hamiltonian
structure. Under the standard definition of the Hamiltonian 
functional $H[p]$ as the Legendre transform of
the Lagrangian $L[v]$, $v=\delta H/\delta p$ and Eq. (11) becomes
\begin{equation}
\frac{\partial p}{\partial t}=
-\frac{\partial}{\partial x}\Big(p\frac{\delta H}{\delta p}\Big)
-p \frac{\partial}{\partial x}\frac{\delta H}{\delta p}.
\label{p_t_Hamiltonian}
\end{equation}
The variational principle $\delta\int{\cal L}dt =0$ for such dynamic
systems has a somewhat unusual form
\begin{equation}
{\cal L}=-\int \sqrt{p}\partial_x^{-1}\partial_t\sqrt{p}\,dx\, -H[p]. 
\end{equation}
For any Hamiltonian, there is the law of conservation
\begin{equation}
\int \sqrt{p}\, dx=S=\mbox{const}.
\label{S}
\end{equation}
The case where $p(x,t)$ is a sign-alternating function
remains questionable.

\section*{Equations for the coefficients}

The substitution of Eqs. (5) and (10) into Eq. (11),
multiplication of the result by $(d/dx)^{-1}P_m(x)$, and 
integration over $x$ yield the infinite system of ordinary 
differential equations
\begin{equation}
\frac{2D_m}{2m+1}\dot a_m=2\lambda\sum_n\Omega_{mn} a_n
+\sum_{n,k}W_{mnk} a_n D_k a_k,
\label{a_t}
\end{equation}
where the elements of the antisymmetric matrix $\Omega$ are
given by the expression
\begin{equation}
\Omega_{mn}=-\Omega_{nm}=-\int_{-1}^{1}P_n(x)
\Big(\frac{d}{dx}\Big)^{-1}P_m(x) dx.
\end{equation}
In view of the properties of the Legendre polynomials,
only the matrix elements between neighboring modes
are nonzero; they are given by the formula
\begin{equation}
\Omega_{m,m-1}=\frac{2}{(2m+1)(2m-1)}.
\end{equation}
The quadratic nonlinearity tensor W is also antisymmetric 
in its first two indices. After integration by parts, it
is reduced to the form
\begin{equation}
W_{mnk}=\Big[\frac{1}{m(m+1)}-\frac{1}{n(n+1)}\Big]J_{mnk},
\end{equation}
\begin{equation}
J_{mnk}=\int_{-1}^{1}(1-x^2)P'_m(x)P'_n(x)P_k(x) dx.
\end{equation}
Here, nonzero elements are obviously only those for
which $m+n+k=2l$, where $l$ is an integer, and $(m+n)\geq k$.
The first nonzero elements are $W_{1,2,1}=4/15$, $W_{1,2,3}=-4/35$,
$W_{1,3,2}=2/7$, $W_{2,3,1}=4/35$, and $W_{2,3,3}=2/105$.

Since $\Omega$ and $W$ are antisymmetric, the dissipation
rate on the optimal cycle is an integral of motion, i.e.,
constant in time:
\begin{equation}
\sum_m\frac{D_m}{2m+1}a_m^2=E=\mbox{const}.
\label{ellipsoid}
\end{equation}
Thus, motion occurs on an ellipsoid in the multidimensional 
phase space.

\begin{figure}
\begin{center}
(a)\epsfig{file=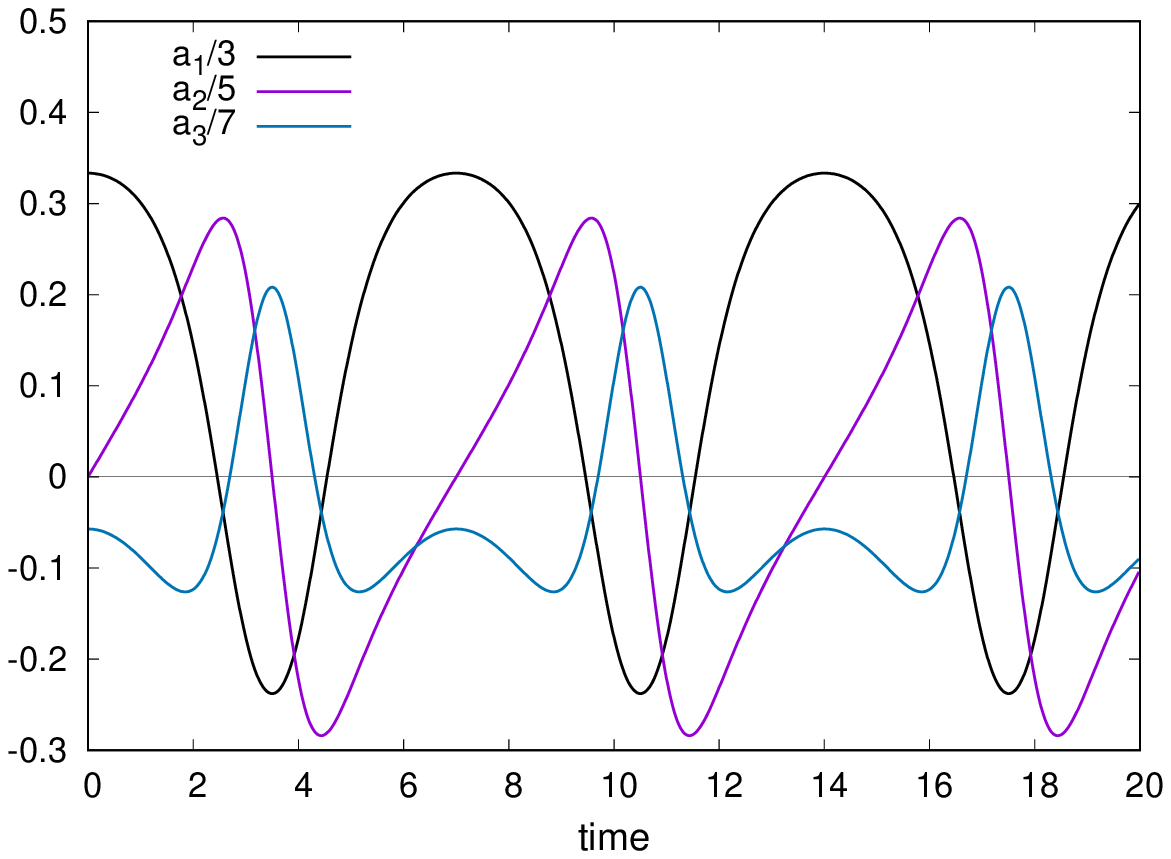, width=72mm}\\
(b)\epsfig{file=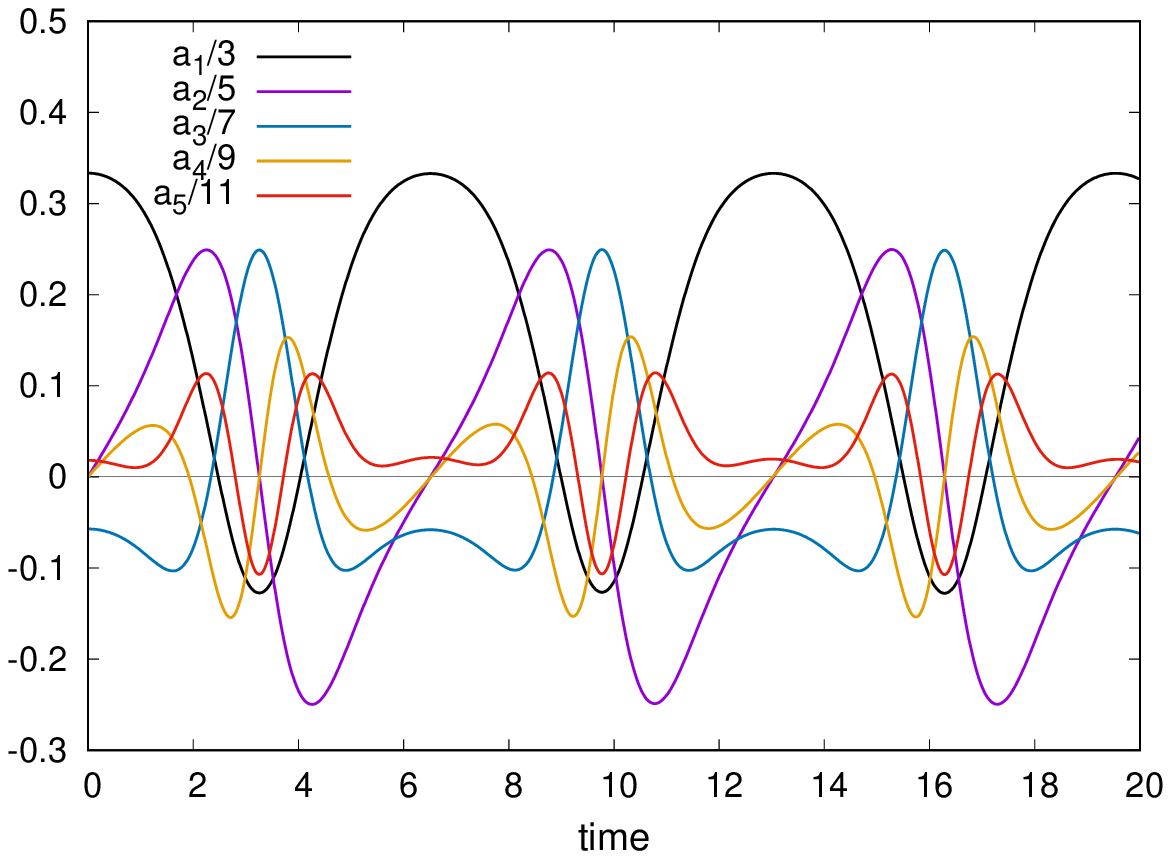, width=72mm}
\end{center}
\caption{(a) Three- and (b) five-mode approximations.}
\label{N3_N5} 
\end{figure}

\section*{Finite-mode approximations}

The numerical results obtained in [16] indicate that
the amplitudes $a_n$ on the optimal solution decrease
exponentially with increasing number $n$. A finite-dimensional 
dynamic system obtained by setting $a_n=0$ 
at $n>N$ in Eqs. (16) is no longer Hamiltonian.
The law of conservation given by Eq. (15) is also ``violated.'' 
However, such approximations can be useful if
one finds their solutions such that highest modes carry
only a small fraction of the total ``energy.'' In this context, 
the cases $N=3$ and $N=5$ are remarkable. The equations 
in the case $N=3$ have the form
\begin{eqnarray}
\frac{2D_1}{3}\dot a_1&=&-\frac{4\lambda}{15}a_2+
\frac{4}{15} a_2 D_1 a_1\nonumber\\
&-&\frac{4}{35} a_2 D_3 a_3+\frac{2}{7} a_3 D_2 a_2,\\
\frac{2D_2}{5}\dot a_2&=&\frac{4\lambda}{15}a_1-
\frac{4\lambda}{35}a_3 
-\frac{4}{15} a_1 D_1 a_1\nonumber\\
&+&\frac{4}{35} a_1 D_3 a_3 +\frac{4}{35} a_3 D_1 a_1+
\frac{2}{105} a_3 D_3 a_3,\\
\frac{2D_3}{7}\dot a_3&=& \frac{4\lambda}{35}a_2-
\frac{2}{7} a_1 D_2 a_2\nonumber\\
&-&\frac{4}{35} a_2 D_1 a_1-\frac{2}{105} a_2 D_3 a_3.
\end{eqnarray}
This system has an additional integral of motion
$F(a_1,a_3)=C$, because the factor $a_2$ is common for all
terms on the right-hand sides of Eqs. (22) and (24).
Thus, the ratio of Eqs. (22) and (24) is free of the variable 
$a_2$ and corresponds to an autonomous linear
inhomogeneous system with a focus singular point.
The phase trajectory is the intersection of the ellipsoid
given by Eq. (21) and a surface from the family
$F(a_1,a_3)=C$ (it is assumed that the focus is beyond the
ellipsoid). An example of the three-mode dynamics is
shown in Fig. 1a. In the case $N=5$, initial conditions
can be easily chosen such that the phase trajectory is
nearly periodic and passes through relatively small $a_4$
and $a_5$ values, as shown in Fig. 1b. This picture is qualitatively 
very similar to Fig. 6 in [16]. However, Fig. 1b
quantitatively presents a different optimal cycle
because the action functional really has many local
minima, as mentioned by the authors of [16].

An algorithm for numerically seeking appropriate
periodic solutions for systems with $N>5$ has not yet
been developed. The trial-and-error method gave no
results for such systems.

It should be emphasized that the problem of optimization 
does not end with solving the dynamic system 
given by Eq. (16). Further, it is necessary to select
solutions on which the displacement is really maximal.
However, some slightly nonoptimal solutions can be
preferable under additional constraints on the properties 
of Lagrangian mappings. For example, if the most
optimal solution dictates overly large displacements of
Lagrangian markers in the angle $\theta$ that are impossible
according to the internal structure of a real squirmer,
solutions with a smaller amplitude can be considered.

\section*{Conclusions}

To summarize, the Eulerian description of the
optimal dynamics of the spherical squirmer has
revealed its noncanonical Hamiltonian structure.
Furthermore, within this description, equations of
motion convenient for numerical solution have been
derived directly in terms of the coefficients of 
expansion of the velocity field on the surface. The found
approximate solutions and their arrangement in the
phase space make it possible to better understand the
behavior of several most important first modes.

A similar approach can also be used for other
axisymmetric squirmers, but the velocity field $v(x,t)$
should be expanded in appropriate sets of functions
depending on the geometrical shape of an object.


\begin{thebibliography}{99}

\bibitem{LR2009} E. Lauga and T. R. Powers, Rep. Prog. Phys. {\bf 72}, 
096601 (2009).

\bibitem{Purcell1977} E. M. Purcell, Am. J. Phys. {\bf 45}, 3 (1977).

\bibitem{NG2004} A. Najafi and R. Golestanian, Phys. Rev. E {\bf 69}, 
062901 (2004);

\bibitem{GA2008} R. Golestanian and A. Ajdari, Phys. Rev. E {\bf 77},
036308 (2008).

\bibitem{ADeSL2009} F. Alouges, A. DeSimone, and A. Lefebvre, 
Eur. Phys. J. E {\bf 28}, 279 (2009).

\bibitem{ZNM2009} R. Zargar, A. Najafi, and M. Miri, Phys. Rev. E {\bf 80},
026308 (2009).

\bibitem{AGK2004} J. E. Avron, O. Gat, and O. Kenneth, 
Phys. Rev. Lett. {\bf 93}, 186001 (2004).

\bibitem{GS2006} E. Gauger and H. Stark, Phys. Rev. E {\bf 74}, 
021907 (2006).

\bibitem{TH2007} D. Tam and A. E. Hosoi, Phys. Rev. Lett. {\bf 98}, 
068105 (2007).

\bibitem{T2015} D. Takagi, Phys. Rev. E {\bf 92}, 023020 (2015).

\bibitem{Lighthill1952} M. J. Lighthill,  
Commun. Pure Appl. Math. {\bf 5}, 109 (1952).

\bibitem{Blake1971} J. R. Blake, J. Fluid Mech. {\bf 46}, 199 (1971).

\bibitem{PL2014} O. S. Pak and E. Lauga, J. Eng. Math. {\bf 88}, 1 (2014).

\bibitem{TWGW2016}
M. Theers, E. Westphal, G. Gompper, and R. G. Winkler, 
Soft Matter {\bf 12}, 7372 (2016).

\bibitem{PA2017} D. Papavassiliou and G. P. Alexander, 
J. Fluid Mech. {\bf 813}, 618 (2017) 

\bibitem{ML2010} S. Michelin and E. Lauga, 
Physics of Fluids {\bf 22}, 111901 (2010).

\end{thebibliography}
\end{document}